\documentclass[12pt]{iopart}

\usepackage{amssymb}
\usepackage{ifpdf}
\usepackage[utf8]{inputenc}
\usepackage{graphicx}
\usepackage{hyperref}
\usepackage{bm}
\usepackage{textcomp}
\usepackage{times}

\newcommand{\URuSi}{URu$_{2}$Si$_2$}

\begin{document}
\title[]{Absence of a static in-plane magnetic moment in the ``hidden-order'' phase of \URuSi}

\author{P Das$^1$, R E Baumbach$^1$, K Huang$^2$, M B Maple$^2$, Y Zhao$^{3,4}$, J S Helton$^3$, J W Lynn$^3$, E D Bauer$^1$ and M Janoschek$^{1,2}$}

\address{$^1$ Condensed Matter and Magnet Science, Los Alamos National Laboratory, Los Alamos, New Mexico 87545, USA}
\address{$^2$ Department of Physics, University of California, San Diego, La Jolla, California 92093, USA}
\address{$^3$ NIST Center for Neutron Research, National Institute of Standards and Technology, Gaithersburg, Maryland 20899, USA}
\address{$^4$ Department of Materials Science and Engineering, University of Maryland, College Park, Maryland 20742, USA}
\ead{mjanoschek@lanl.gov}

\begin{abstract}
We have carried out a careful magnetic neutron scattering study of the heavy fermion compound \URuSi\ to probe the possible existence of a small magnetic moment parallel to tetragonal basal plane in the ``hidden-order''  phase. This small in-plane component of the magnetic moment on the uranium sites $S_\parallel$ has been postulated by two recent models (rank-5 superspin/hastatic order) aiming to explain the hidden-order phase, in addition to the well-known out-of-plane component $S_\perp$~$\approx$~0.01-0.04 $\mu_B$/U. In order to separate $S_\parallel$ and $S_\perp$ we take advantage of the condition that for magnetic neutron scattering only the components of the magnetic structure that are perpendicular to the scattering vector $Q$ contribute to the magnetic scattering. We find no evidence for an in-plane magnetic moment $S_\parallel$. Based on the statistics of our measurement, we establish that the upper experimental limit for the size of any possible in-plane component is $S^{\rm{max}}_\parallel$~$\leq$~1$\cdot$10$^{-3}$~$\mu_B$/U.
\end{abstract}

\submitto{\NJP}
\maketitle

\section{Introduction}

In compounds that contain transition metal, lanthanide, or actinide ions with partially filled $d$- or $f$ -electron shells, the strong electronic correlations originating in the hybridization of localized $d$- or $f$ -electron and conduction electron states often leads to the emergence of new electronic  ground states such as heavy fermion metals, complex magnetic order, quadrupolar order, non-Fermi-liquid (NFL) behavior,  and unconventional superconductivity (SC) \cite{Maple:10}. The search for and the understanding of novel electronic ground states thus is an important research direction in the study of strongly correlated electron phenomena.

A prime example of such emergent behavior is the ``hidden-order'' (HO) phase in the heavy fermion compound \URuSi\ that occurs below $T_0$ = 17.5 K and coexists with SC below $T_c$ = 1.5 K~\cite{Palstra:85, Maple:86, Schlablitz:86}. Neutron scattering experiments demonstrate the presence of a small antiferromagnetic moment $S_{\perp}$~$\approx$~0.01-0.04 $\mu_B$/U perpendicular to the tetragonal basal plane in the HO phase that is, however, much too small to account for the entropy of $\approx$ 0.2Rln(2) associated with the observed specific-heat anomaly~\cite{Broholm:87, Amitsuka:07, Niklowitz:2010}. Recently, more detailed neutron scattering studies have, however, put forth the view that the magnetic structure in the HO phase is induced by strain~\cite{Amitsuka:07, Niklowitz:2010} and is due to a small amount of the neighboring large moment antiferromagnetic (LMAFM) phase that emerges at critical pressures $P_c$~$\geq$~0.5-1.5~GPa~\cite{Amitsuka:99}.

This led to the terminology HO~\cite{Luethi:93} to allude to the unknown identity of the corresponding order parameter (OP) that has eluded identification for almost three decades. Notably, the search for the OP of the HO phase has attracted an enormous amount of attention, and over the last few decades the full arsenal of experimental methods has been employed in the effort to unravel this notorious phase. This concentrated experimental effort has established that the presence of HO is reflected in many details of the complex electronic structure of \URuSi~\cite{Mydosh:11}.

As originally inferred from the specific heat a charge gap of $\Delta$~$\approx$~11~meV opens over about 40\% in the Fermi surface~\cite{Palstra:85,Maple:86}. This is also evident from the large jump observed in the Hall coefficient $R_H$ at $T_0$ that indicates a reduction of charge carrier concentration $n$ from 0.10 holes per U atom in the paramagnetic state to 0.02 in the HO phase~\cite{Schoenes:87, Oh:07, Kasahara:07} and measurements of the optical conductivity~\cite{Bonn:88}. Inelastic neutron scattering has further revealed that a spin gap opens simultaneously with the charge gap. Here a spin gap is observed both at the commensurate wave vector $\bm{Q}_0 = (1, 0, 0)$, as well as the incommensurate wave vector $\bm{Q}_1 = (0.4, 0, 0)$, where the values of the corresponding gaps are 2 and 4~meV, respectively~\cite{Broholm:91, Wiebe:07}. Above $T_0$ the $\bm{Q}_0$ mode transforms to weak quasielastic spin fluctuations, where Wiebe {\it et al.} have shown that the gapping of these spin fluctuations accounts for the loss of entropy at the HO transition \cite{Wiebe:07}. In contrast, the $\bm{Q}_1$ mode is due to itinerant-like spin excitations that are related to the heavy electronic quasiparticles that form below a coherence temperature $T^\ast$~$\approx$~70~K.

Further details have been revealed by investigations of the Fermi surface (FS) of \URuSi. Here quantum oscillation measurements~\cite{Ohkuni:99, Altarawneh:11} have demonstrated that the FS in the HO phase of \URuSi\ is mostly dominated by small closed pockets which is again in agreement with the partial gaping of the FS. Moreover, measurements including angle-resolved phohto emission spectroscopy (ARPES)~\cite{Santander-Syro:09}, scanning-tunneling microscopy (STM)~\cite{Schmidt:10, Aynajian:10}, and point-contact spectroscopy (PCS)~\cite{Rodrigo:97} have revealed that the electronic structure of \URuSi\ is reorganized below $T_0$ where a heavy quasiparticle band shifts below the Fermi level, and the crossing with a light hole-like band at $Q^\ast=\pm0.3\pi/a$ leads to the formation of a hybridization gap $\Delta_{Q^\ast}$~=~5~meV.

In parallel with these extensive experimental efforts the HO problem has also motivated a wide range of theoretical studies, in particular because many of the electronic signatures of the HO phase are also relevant more broadly for strongly correlated electron systems in general. In turn a multitude of models to explain the HO phase and its elusive nature have been proposed (see Ref.~\cite{Mydosh:11} and references therein). However, to date none of the models could be confirmed, often because the nature of the corresponding complex OP cannot be easily verified by means of current experimental methods.

Recent magnetic torque experiments suggest that the HO phase spontaneously breaks the rotational symmetry of the crystal in the [110] direction~\cite{Okazaki:11}. In combination with the absence of lattice distortions through the HO transition~\cite{Kernavanois:99,Kuwahara:97},  this suggests that the broken symmetry is solely of electronic origin. Interestingly, this new experimental constraint rules out many previously proposed models for the HO phase (see e.g. Ref.~\cite{Mydosh:11}), and has induced a flurry of new theoretical proposals, such as dynamical symmetry breaking~\cite{Oppeneer:10, Oppeneer:11}, staggered spin-orbit coupling order ~\cite{Das:12}, spin-nematic states~\cite{Fujimoto:11}, a rank-5 superspin ~\cite{Rau:12}, and hastatic order ~\cite{Chandra:13}. Here we focus on the latter two, which both propose a non-zero magnetic moment $S_{\parallel}$ in the tetragonal basal plane -- a prediction that can be experimentally tested via magnetic neutron diffraction.

The OP proposed by Rau \emph{et al.}~\cite{Rau:12} is a rank-5 E type spin density wave between the $5f$ crystal field doublets $\Gamma_7^{(1)}$ and $\Gamma_7^{(2)}$. This candidate for the HO OP is based on a tight binding model for the itinerant heavy quasiparticles in \URuSi. The proposed OP breaks both time reversal symmetry and the lattice point group symmetry $D_{4h}$, consistent with the torque magnetometry results. For this OP a small antiferromagnetic magnetic moment $S_{\parallel}$ in the tetragonal basal plane oriented along the [110] direction ordered at a wave vector (0,0,1) is expected due to second-order correlations.

Chandra \emph{et al.}~\cite{Chandra:13} have based their proposal for the HO OP on the Ising-like nature of local $5f$ moments in \URuSi~\cite{Palstra:85}, as well as on the recent observation that the quasiparticles in the HO phase exhibit a giant Ising anisotropy \cite{Altarawneh:11, Altarawneh:12}. They demonstrate that such Ising quasiparticles result from a spinor order parameter that breaks double time-reversal symmetry, mixing states of integer and half-integer spin by hybridizing the conduction electrons with Ising $5f^2$ states of the uranium atoms. This OP accounts for the large specific heat jump and the torque magnetometry results. Similar to the proposal by Rau \emph{et al.}~\cite{Rau:12} a small in-plane magnetic moment $S_{\parallel}$ is predicted. Estimates from this theory suggest that $S_{\parallel}$ is of the order of 0.015 $\mu_B$/U, which would be detectable in a sensitive neutron diffraction experiment.

To date such an in-plane magnetic moment in the HO phase of \URuSi\ has not been observed. Establishing an experimental limit for the size of $S_{\parallel}$ is therefore highly desirable in order to guide current and future theoretical efforts in disentangling the nature of the HO phase. Here we show by means of a carefully designed neutron diffraction study that within the detection limit of state-of-the-art neutron scattering no static in-plane magnetic moment exists within the HO phase.

\section{Experimental Details}

The sample used for this work is a 7 g single crystal of \URuSi\ synthesized via the Czochralski technique using a tetra-arc furnace. The sample has been characterized by both x-ray and neutron diffraction measurements. High sample quality is demonstrated by a residual resistance ratio of $\approx$10 within the tetragonal plane that is identical to other neutron scattering studies of high quality single crystals ~\cite{Niklowitz:2010}.

\begin{figure}[t]
  \includegraphics[width=0.8\linewidth]{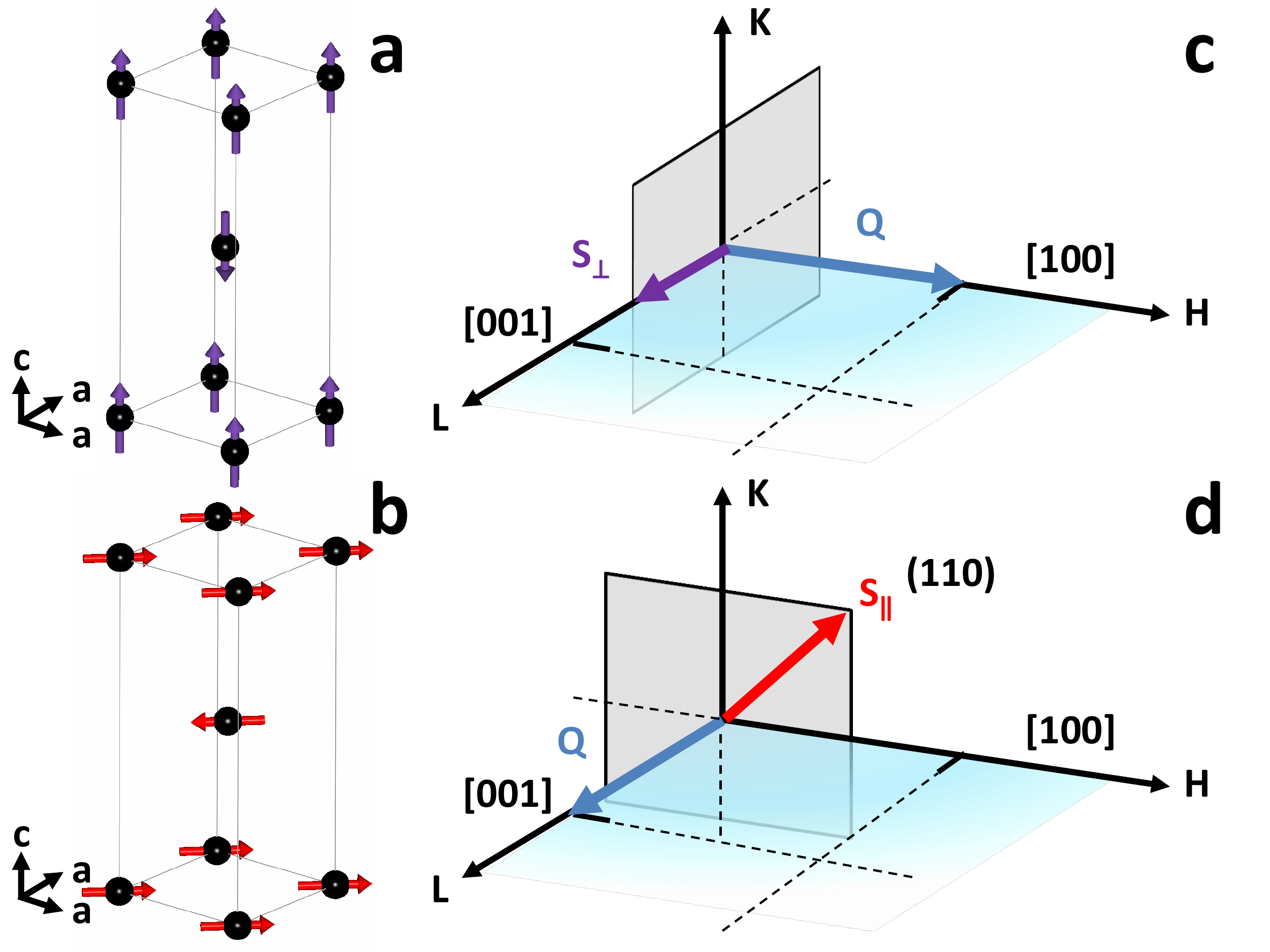}
  \caption{The magnetic structure associated with the hidden-order phase of \URuSi\ is illustrated (a,b) together with the geometry used in our neutron diffraction experiment to disentangle the (c) \emph{known} and (d)~\emph{theoretically predicted} components of the magnetic moments. For clarity only the magnetic uranium ions are shown. Panel (a) shows the \emph{known magnetic structure as reported in the literature} ~\cite{Broholm:87, Amitsuka:07, Niklowitz:2010} with the magnetic moments on the two uranium sites perpendicular to the basal plane ($S_{\perp}$). In (b) we show the \emph{theoretically predicted additional component} $S_{\parallel}$ that lies within the tetragonal basal plane and is directed along the (110) direction ~\cite{Rau:12, Chandra:13}. In (c) and (d) we show that our sample was oriented so that only [H0L] reflections were accessible (cf. blue shaded plane). Note that the coordinate frames in (c,d) are rotated 90$^\circ$ around the crystallographic (100)-axis with respect to (a,b). In magnetic neutron scattering only components of the magnetic structure that are perpendicular to the scattering vector $\bm{Q}$ contribute to the scattering (see (a,b) and text for details). (c) Scattering geometry to determine $S_{\perp}$ (purple arrow) by probing the [100] magnetic reflection associated with scattering from magnetic moments in the [0KL] plane (light grey). (d) Scattering geometry to determine $S_{\parallel}$ (red arrow) by probing the [001] magnetic reflection associated with scattering from magnetic moments in the [HK0] plane (light grey).}\label{fig:magscattering}
\end{figure}

\begin{figure}[t]
  \includegraphics[width=0.9\linewidth]{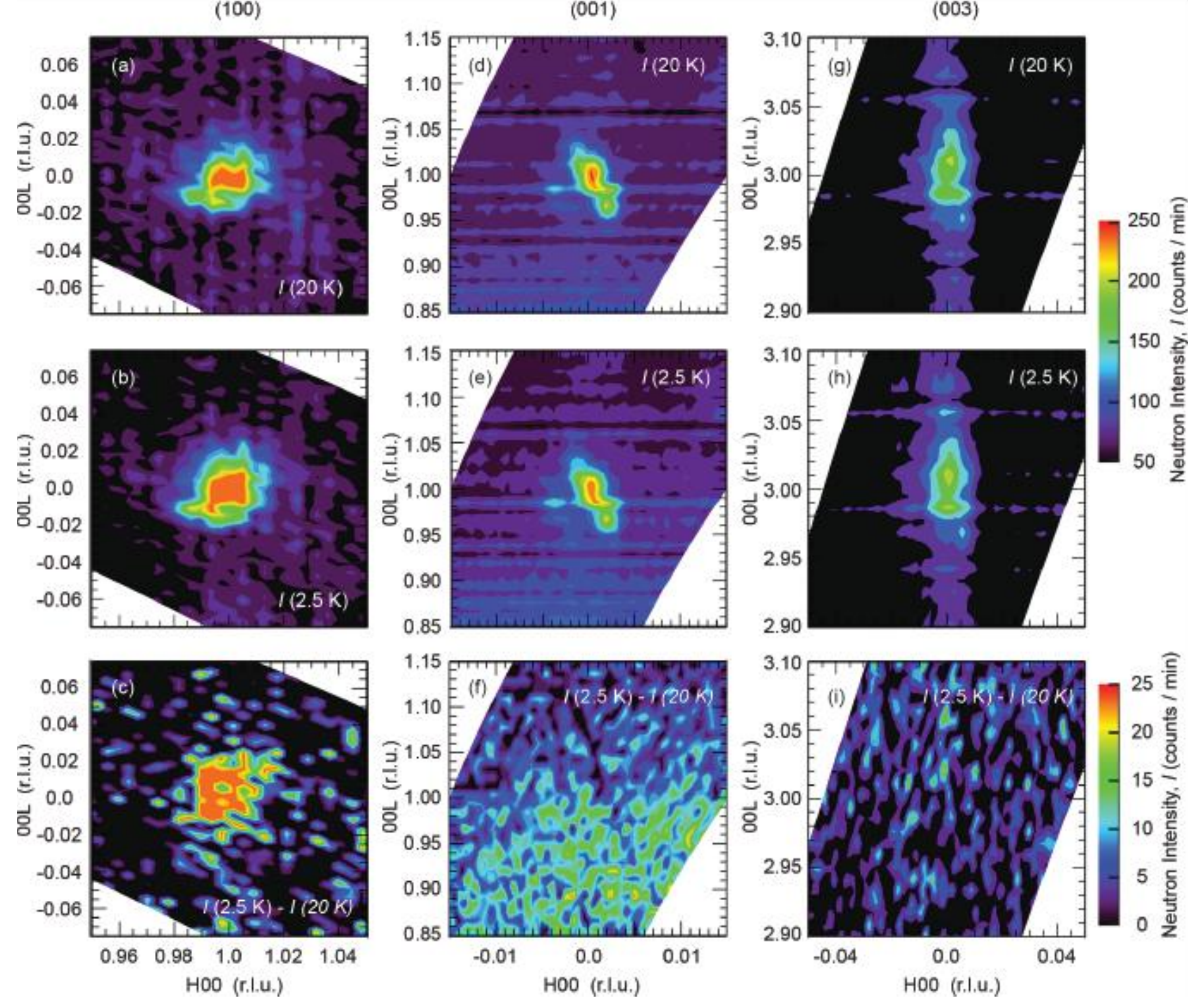}
  \caption{Neutron scattering intensity is observed in maps recorded around the [100](a,b), [001](d,e) and [003](g,h) reciprocal space positions for temperatures $T$~$=$~2.5~K (hidden order state) and  $T$~$=$~20~K (paramagnetic state), respectively. The measurement configuration for the data shown in (a,b) corresponds to Fig.~\ref{fig:magscattering}(c), and therefore probes mostly the component of the magnetic moment perpendicular to the tetragonal basal plane $S_\perp$, whereas (d,e,g,h) were measured in the configuration illustrated in Fig.~\ref{fig:magscattering}(d) and investigate the existence of the in-plane component $S_\parallel$. For all three reciprocal space positions measured, the intensity observed in the paramagnetic state is due to contamination from higher order scattering (see text). In (c),(f) and (i) we show maps where the intensity of $T$~$=$~20~K was subtracted from the $T$~$=$~2.5~K data set to remove the higher order scattering for the [100], [001], and [003] positions, respectively. The data in (a-c) were recorded by counting 1 min per point. In contrast data in (d-i) was obtained by counting 12 min per point.}\label{fig:maps}
\end{figure}

Our measurements were carried out on the BT-7 thermal triple axis spectrometer at the NIST Center for Neutron Research \cite{Lynn:2012}. BT-7 was operated in elastic mode with the wavelength $\lambda$~$=$~2.36~\AA. To reduce contamination due to higher-order Bragg scattering at the monochromator PG filters both up and down stream of the sample position were employed. The instrument was optimized for high intensity in order to search for the postulated in-plane magnetic moment and used a position-sensitive detector (PSD) without the analyzer. BT-7 was equipped with an 80' Soller collimator in front of the sample position and an 80' radial collimator in front of the PSD.

\section{Separating in-plane and out-of-plane magnetic moments}

\begin{figure}[t]
  \includegraphics[width=0.99\linewidth]{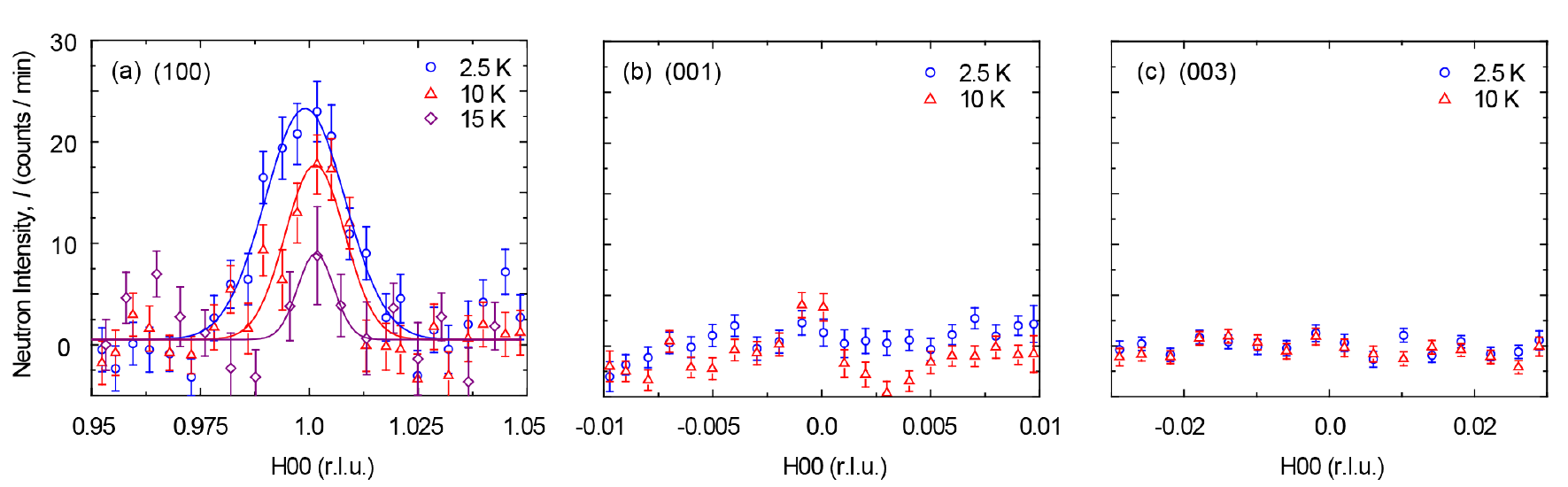}
  \caption{Neutron scattering intensity observed in $\bm{Q}$-scans through the (a) [100], (b)[001], and (c)[003] reciprocal positions for various temperatures. The $\bm{Q}$-scans have been extracted from the maps such as the ones in Fig.~\ref{fig:maps}. (a) Integrated intensities have been obtained by fitting the observed magnetic peaks with Gaussian profiles and a background. Uncertainties, where indicated are statistical in origin and represent one standard deviation.}\label{fig:qscans}
\end{figure}

In order to separate the postulated in-plane ($S_{\parallel}$) from the well-known out-of-plane ($S_{\perp}$) component of the weak magnetic moment we have exploited the condition that the magnetic scattering intensity depends on the mutual direction between the magnetic moments and the scattering vector $\bm{Q}$. The magnetic neutron scattering cross-section is proportional to $\sum_{\alpha,\beta}(\delta_{\alpha\beta}-\hat{Q}_{\alpha}\hat{Q}_{\beta})M_{\bm{Q}}^{\alpha\dagger}M_{\bm{Q}}^{\beta}$, where $\hat{Q}$ and $\bm{M}_{\bm{Q}}$ are a unit vector parallel to the scattering vector $\bm{Q}$ and the magnetic structure factor, respectively.  $\alpha$ and $\beta$ describe their components with $\alpha,\beta = x,y,z$. The magnetic structure factor is the Fourier transform of the local magnetization density $\bm{M}(\bm{r})$ and can be written in terms of magnetic Fourier components \cite{Izyumov:91} that for \URuSi\ take the form $\bm{M}_{\bm{Q}}=\sum_d g_d F_d(\bm{Q})\exp(-W_d(\bm{Q}))\exp(i\bm{Q}\cdot\bm{d})\bm{S}^{\bm{Q}}_d$. Here $\bm{d}$ describes the position of the $d$th magnetic ion in the unit cell. $g_d$, $F_d(\bm{Q})$, and $\exp(-W_d(\bm{Q}))$ are the Land\'{e} $g$-factor, the magnetic form factor and the Debye-Waller factor for the $d$th ion, respectively.

\URuSi\ crystallizes in the space group $I 4/m m m$, and the antiferromagnetic order found in the HO state is described by the two uranium ions (0,0,0) and  (1/2, 1/2 1/2) that exhibit antiparallel magnetic moments $\bm{S}^{\bm{Q}}_1 = (S_{\parallel,x}, S_{\parallel,y}, S_\perp)$ and $\bm{S}^{\bm{Q}}_2 = (-S_{\parallel,x}, -S_{\parallel,y}, -S_\perp)$, where $S_{\parallel} = \sqrt{S_{\parallel,x}^2+S_{\parallel,y}^2}$. We denote the known component perpendicular to the tetragonal basal plane with $S_\perp$ (Fig.~\ref{fig:magscattering}(a)), and the postulated component within the plane as $S_\parallel$ ( Fig.~\ref{fig:magscattering}(b)).

Inspecting the $(\delta_{\alpha\beta}-\hat{Q}_{\alpha}\hat{Q}_{\beta})$ term in the magnetic cross-section it is clear that only components of the magnetic structure factor that are perpendicular to $\bm{Q}$ are visible for the current scattering geometry as illustrated in Fig.~\ref{fig:magscattering}. Our sample was oriented in such a way that only [H0L] Bragg reflections were accessible. In these experiments we probed the components of the magnetic moments in two configurations. By measuring the intensity of the [100] magnetic Bragg reflection, only components parallel to the [0KL] plane (equivalent to the real space $ac$-plane) may be observed thus allowing us to probe the known component $S_\perp$ (Fig.~\ref{fig:magscattering}(c)). Similarly, the existence of $S_\parallel$ was investigated by measuring at the [001] position, which only generates magnetic intensity from spin components within the tetragonal basal plane (Fig.~\ref{fig:magscattering}(d)). Furthermore, we have also measured the [003] reciprocal position, for which the geometry is identical to the [001] position, and therefore was also used to probe $S_\parallel$. We note that both the [100], [001], [003] reflections are forbidden structural reflections and all scattering observed at these positions will be purely magnetic.

\section{Results}

Fig.~\ref{fig:maps} shows maps of the observed neutron intensity as measured around the [100], [001] and [003] Bragg reflections, respectively. The maps were obtained by rotating the sample around the axis perpendicular to the [H0K] scattering plane used in our experiment. For the [100] reflection, which probes mostly $S_\perp$, there is a difference in neutron intensity for the data taken at $T$~$=$~2.5~K in the HO phase (Fig.~\ref{fig:maps}(a)), and  $T$~$=$~20~K above $T_0$ (Fig.~\ref{fig:maps}(b)). Even in the 20~K data set a clear peak is visible which is due to higher-order scattering from the [002] reflection which is the most intense nuclear Bragg reflection. This was verified at the end of our experiment by inserting a third PG filter after the sample position, which in turn eliminated the higher-order scattering completely. To isolate the magnetic scattering from the temperature-independent higher order scattering, we additionally show a map that was obtained by subtracting the 20~K from the 2.5~K data set (Fig.~\ref{fig:maps}(c)). As seen in this difference plot, there is a clear magnetic response corresponding to the previously observed magnetic structure with the magnetic moment parallel to the $c$-axis arises below $T_0$.

In contrast, for the [001] reciprocal space position that purely probes $S_\parallel$, there is no visible difference between measurements at $T$~$=$~2.5~K (Fig.~\ref{fig:maps}(d)) and ~20~K (Fig.~\ref{fig:maps}(e)). This is also borne out in the difference data (Fig.~\ref{fig:maps}(f)) indicating that within the detection limit of neutron diffraction $S_\parallel$~$\approx$~0. Similar to the [100] position the temperature-independent intensity is due to higher order nuclear scattering. We note that the [001] reciprocal space position in our setup with $\lambda$~$=$~2.36~\AA~ is near to the unscattered neutron beam, e.g. the scattering angle for [001] is 2$\theta$~$=$~14.2 degrees, and thus is affected by an increased background. Therefore, we have carried out additional measurements around the [003] reciprocal space position that also is purely sensitive to $S_\parallel$. While measurements at the [003] position are less sensitive to magnetic scattering compared to [001] because the magnetic form factor for U$^{4+}$ ions is almost 30\% reduced, the background is significantly improved at [003]. The corresponding maps for [003] for $T$~$=$~2.5~K and ~20~K are shown in Figs.~\ref{fig:maps}(g) and (h), respectively, and the difference map is shown in Fig.~\ref{fig:maps}(i). Just as for the [001] position there is no additional magnetic intensity observed at the [003] reflection, confirming that $S_\parallel$~$\approx$~0.

In order to provide an experimental upper limit for the size of $S_\parallel$ we plot $\bm{Q}$-scans through the [100], [001], and [003] positions shown in Fig.~\ref{fig:qscans} that have been extracted from the maps such as the ones in Fig.~\ref{fig:maps}. In addition, we have measured more maps in the HO phase at [100] at $T$~$=$~10 and 15~K, and at [001] and [003] at $T$~$=$~10~K, and the corresponding extracted $\bm{Q}$-scans are also plotted in Fig.~\ref{fig:qscans}. For all shown $\bm{Q}$-scans the data recorded at $T$~$=$~20~K was subtracted to remove the higher-order nuclear scattering.

\begin{figure}[t]
  \includegraphics[width=0.6\linewidth]{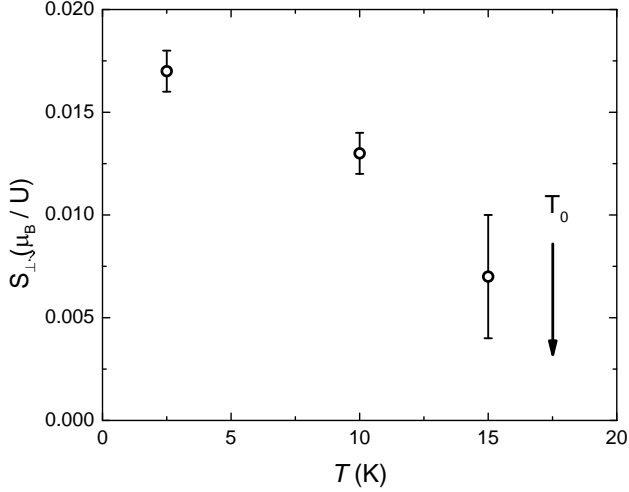}
  \caption{The temperature dependence of the magnetic moment perpendicular to the tetragonal plane $S_\perp$ as measured on the [100] magnetic Bragg reflection is shown vs. temperature $T$. $T_0$ denotes the transition temperature to the hidden-order phase.}\label{fig:temp}
\end{figure}

$\bm{Q}$-scans through the [100] (cf. Fig.~\ref{fig:qscans}(a)) position show clear magnetic peaks arising in the HO phase. The integrated intensity of the [001] magnetic Bragg reflection was determined by means of fits with Gaussian profiles. The corresponding magnetic moment perpendicular to the tetragonal basal plane was calculated from the integrated intensity by calibrating with the integrated intensity of the [006] nuclear Bragg reflection. This reflection was the weakest nuclear Bragg reflection accessible in our experiment and was chosen to avoid the problem of extinction. Using this method we found that $S_\perp$~$=$~0.016(1)~$\mu_B$/U in agreement with the values reported in literature previously~\cite{Broholm:87, Amitsuka:07, Niklowitz:2010}. The temperature dependence of $S_\perp$ is shown in Fig.~\ref{fig:temp}.

As demonstrated in Figs.~\ref{fig:qscans}(b) and (c) the $\bm{Q}$-scans through the [001] and [003] positions show no peak at all, and no temperature dependence, clearly indicating the absence of any moment in the tetragonal basal plane. We note, that all measurements on the [100] position have been performed with count rates of 1 min per point, whereas the measurements around [001] and [003] have been carried out with 12 min per point,
in turn resulting in 3.5 times better statistics. This allows us to define a simple experimental upper limit for  $S_\parallel$. As illustrated in Fig.~\ref{fig:temp} the magnitude of the size of magnetic moment perpendicular to the basal plane decreases to $S_\perp$~$=$~0.007(3)~$\mu_B$/U at $T$~$=$~15~K, which is just below $T_0$. Notably, from the $\bm{Q}$-scans through the [100] magnetic Bragg reflection at $T$~$=$~15~K shown in Fig.~\ref{fig:qscans}(a) it is clear that $S_\perp(15$~ K$)$ is at the detection limit of our experiments for scans performed with count rates of 1 min per point.

Using the fact that the $\bm{Q}$-scans around the [001] and [003] reciprocal space positions have been carried out with 3.5 better statistics, we thus can define a conservative detection limit for the in-plane component $S^{\rm{conservative}}_\parallel$~$\leq$~$S_\perp(15$~ K$)$/3.5~$=$~2$\cdot$10$^{-3}$~$\mu_B$/U. However, using both the magnetic form factor dependence for the reciprocal space positions investigated and the known temperature dependence of the known component of the magnetic moment $S_\perp$, we may use the full statistics of the four $\bm{Q}$-scans shown in Fig.~\ref{fig:qscans}(b) and (c) to define a more accurate detection limit.

Considering the magnetic form factor for U$^{4+}$ ions on all three measured reciprocal space positions [100], [001] and [003], we find that $F_d([001])/F_d([100])$~$=$~1.20 and $F_d([003])/F_d([100])$~$=$~0.86. Furthermore, inspecting Fig.~\ref{fig:temp} we get $S_\perp ($10~K$)$/$S_\perp ($2.5~K$)$~$=$~0.8. Therefore, the full statistics of the four $\bm{Q}$-scans performed at the [001] and [003] positions is a factor $\sqrt{12\cdot\left\{\left[(F_d([001])/F_d([100]))^2+(F_d([003])/F_d([100]))^2\right]\cdot\left[1+(\frac{S_\perp (\rm{10~K})}{S_\perp (\rm{2.5~K})})^2\right]\right\}}$~$=$~6.5 better than for the $\bm{Q}$-scan carried out at the [100] position. Here we have used that the magnetic neutron intensity is proportional to the square of both the magnetic moment and the magnetic form factor. Consequently $\bm{Q}$-scans carried out around the [001] and [003] reciprocal space positions with 6.5 better statistics would be able to detect a magnetic moment $S^{\rm{max}}_\parallel$~$\leq$~$S_\perp(15$~ K$)$/6.5~$=$~1$\cdot$10$^{-3}$~$\mu_B$/U.

Finally, our results additionally enable us to establish an upper limit for the magnetic correlation length $\xi$ between magnetic moments $S_\parallel$ in the tetragonal basal plane. The width of a magnetic Bragg in a neutron scattering experiment is a direct measure for the inverse magnetic correlation length $\kappa=1/\xi$. The region of the reciprocal space that we have investigated around the [001] and [003] directions (cf. Fig.~\ref{fig:maps}) allows us to probe maximum inverse correlation lengths  $\kappa_a$~$=$~0.2~\AA$^{-1}$ and $\kappa_c$~$=$~0.12~\AA$^{-1}$ along the crystallographic $a$ and $c$ directions, respectively. Because we have not observed magnetic Bragg peaks over these inverse length scales any magnetic correlations in \URuSi\ would have larger inverse correlation length, i.e. the associated magnetic peak would be much broader than the large portion of reciprocal space investigated in this experiment. This suggests that any magnetic correlations between in-plane magnetic moments $S_\parallel$ must develop over correlations lengths smaller than $\xi_a^{max}=1/\kappa_a$~$=$~8.2~\AA~ and $\xi_c^{max}=1/\kappa_c$~$=$~5.1~\AA, which corresponds to approximately 2 and 0.5 lattice spacings for the $a$ and $c$ axes, respectively. This clearly establishes that any sort of magnetic correlations for magnetic moments parallel to the tetragonal basal plane must be very short-range in nature, and that no long-range magnetic order with in-plane moments develops in the HO phase of \URuSi.

\section{Discussion and Conclusion}

We note that the absence of a static in-plane component of the magnetic moment $S_\parallel$ may be explained in terms of a fluctuating in-plane component. This possibility was not investigated in detail in this work, because current models suggest a static magnetic moment \cite{Rau:12, Chandra:13}. However, we note that because our experiment was carried out in the diffraction mode of BT-7 where the intensities are integrated over all final neutron energies after the scattering process at the sample, a fluctuating moment $S_\parallel$ should have still led to some increased intensity at the [003] reciprocal space position if both the moment and the lifetime of the fluctuations were large enough. If the lifetime of the fluctuations were short or the fluctuating moment was very weak, only an energy-resolved neutron scattering experiment would be able to detect an in-plane fluctuating moment. This possibilities will be investigated in future work.

We briefly discuss the possibility of the parasitic character of the well-known out-of-plane component $S_\perp$ of the magnetic moment. There are reports that the tiny magnetic moment $S_\perp$ is induced by strain that causes a small amount of the neighboring LMAFM phase to coexist with the HO phase \cite{Amitsuka:99, Niklowitz:2010} . The presence of this parasitic phase would decrease the volume fraction of the HO phase, and therefore the size of a possible in-plane component $S_\parallel$. However, comparing the measured moment in the LMAFM phase of 0.5 $\mu_B$/U with the known size of $S_\perp$~$\approx$~0.01-0.04 $\mu_B$/U in the HO phase, one estimates that the parasitic LMAFM phase would be less than 8\%. For high quality samples such as the one used for this study it should be even smaller. Therefore, we conclude that a possible parasitic character of $S_\perp$ does not influence the upper bound for $S_\parallel$ significantly. Similarly, we note that our result is not influenced by the mosaic of our sample. In fact, a sample exhibiting mosaicity increases the acceptance of a typically divergent neutron beam. We therefore argue that a small sample mosaic typical for a high quality sample is useful for detecting small magnetic moments.

We also consider the influence of magnetic domains on the limit provided for $S_\parallel$. \URuSi\ crystallizes in the tetragonal space group $I4/mmm$ that does not contain any symmetry operations which may mix in-plane and out-of-plane components of the magnetic moment, and therefore, we limit our discussion to symmetry operations that transform the in-plane components $(S_{\parallel,x}, S_{\parallel,y}, 0)$ of the magnetic moment. Here, the four-fold rotation symmetry of the tetragonal basal plane will lead to four configuration domains for which the magnetic moments will be directed along $(S_{\parallel,x}, S_{\parallel,y}, 0)$, $(-S_{\parallel,x}, S_{\parallel,y}, 0)$, $(S_{\parallel,x}, -S_{\parallel,y}, 0)$, and $(-S_{\parallel,x}, -S_{\parallel,y}, 0)$, respectively. However, the selected measurement geometry (cf. Fig.~\ref{fig:magscattering}(d)) probes all four domains simultaneously, and the total magnetic intensity at the probed reciprocal space positions [001] and [003] is the sum of the individual intensities for each domain. Thus, the presence of magnetic domains does not influence our result.

In conclusion our measurements provide an experimental upper limit for the size of the theoretically predicted component $S_\parallel$ of the antiferromagnetically ordered static magnetic moments in the HO phase of the heavy fermion system \URuSi. The established experimental limit $S^{\rm{max}}_\parallel$~$\leq$~1$\cdot$10$^{-3}$~$\mu_B$/U is more than an order of magnitude smaller than the value $S^{\rm{theo}}_\parallel$~$=$~0.015~$\mu_B$/U suggested by theory \cite{Chandra:13}. This new limit is therefore an important constraint for present \cite{Rau:12, Chandra:13} and future theories that aim to model the OP of the notorious HO phase of \URuSi.

In particular, it is interesting to discuss our results with respect to the magnetic torque experiments that have been interpreted in such a way that the HO phase spontaneously breaks the rotational symmetry of the crystal in the basal plane~\cite{Okazaki:11}. Here it has been suggested that the line-broadening of the NMR signal observed in \URuSi\ for magnetic fields parallel to the basal tetragonal plane \cite{Takagi:07} is due to a non-zero but tiny susceptibility $\chi_{[110]}$ that also breaks time-reversal symmetry. Our measurements clearly demonstrate that long-range magnetic order with magnetic moments parallel to the basal tetragonal plane is absent in \URuSi\ and that any static contribution to $\chi_{[110]}$ is zero. This suggest that both the proposed broken four-fold symmetry in the tetragonal plane and the putative broken time-reversal symmetry are not due to the presence of a magnetic dipole moment in the basal plane. Based on more recent NMR measurements and symmetry analysis it has been argued that the line-broadening and the associated broken time-reversal symmetry may be explained by the presence of magnetic multipoles such as octupoles or triakontadipoles in combination with domains or disorder~\cite{Takagi:12}. In agreement with our study this NMR study by Takagi {\it et al.} also concludes that the in-plane magnetic dipole moment $S_\parallel$ is zero~\cite{Takagi:12}. Finally, the broken rotational symmetry of the tetragonal basal plane may be explained via an electronic nematic state as proposed in Ref.~\cite{Okazaki:11}, however more detailed investigations of the electronic structure of \URuSi\ are required to resolve this issue.

\ack
We thank Filip Ronning and Premala Chandra for useful discussions. Sample synthesis and characterization at UCSD were funded by the U.S. DOE under Grant No. DE FG02-04ER46105. Work at Los Alamos National Laboratory (LANL) was performed under the auspices of the U.S. DOE, OBES, Division of Materials Sciences and Engineering and funded in part by the LANL Directed Research and Development program. M. J. acknowledges financial support through the Alexander von Humboldt foundation.

\section*{References}

\bibliographystyle{unsrt}

\begin{thebibliography}{10}

\bibitem{Maple:10}
Maple M~B, Baumbach R~E, Butch N~P, Hamlin J~J and Janoschek M 2010 Low Temp. Phys. {\bf 161}  4

\bibitem{Palstra:85}
Palstra T~T~M, Menovsky A~A, van~den Berg J, Dirkmaat A~J, Kes P~H, Nieuwenhuys G~J and Mydosh J~A 1985 Phys. Rev. Lett. {\bf 55} 2727

\bibitem{Maple:86}
Maple M~B, Chen J~W, Dalichaouch Y, Kohara T, Rossel C, Torikachvili M~S, McElfresh M~W and Thompson J~D 1986 Phys. Rev. Lett. {\bf 56} 185

\bibitem{Schlablitz:86}
Schlabitz W, Baumann J, Pollit B, Rauchschwalbe U, Mayer H~M, Ahlheim U and Bredl C~D 1986 Z. Phys. B: Condens. Matter {\bf 62} 171

\bibitem{Broholm:87}
Broholm C, Kjems J~K, Buyers W~J~L, Matthews P, Palstra T~T~M, Menovsky A~A and Mydosh J~A 1987 Phys. Rev. Lett. {\bf 58} 1467.

\bibitem{Amitsuka:07}
Amitsuka H, Matsuda K, Kawasaki I, Tenyaa K, Yokoyama M, Sekine C, Tateiwa N, Kobayashi T C, Kawarazaki S and Yoshizawa H 2007 Jour. Mag. Mag. Mater. {\bf 310} 214

\bibitem{Niklowitz:2010}
Niklowitz P~G, Pfleiderer C, Keller T, Vojta M, Huang Y~-K and Mydosh J~A 2010 Phys. Rev. Lett. {\bf 104} 106406

\bibitem{Amitsuka:99}
Amitsuka H, Sato M, Metoki N, Yokoyama M, Kuwahara K, Sakakibara T, Morimoto H, Kawarazaki S, Miyako Y and Mydosh J~A 1999 Phys. Rev. Lett. {\bf  83} 5114

\bibitem{Luethi:93}
Luethi B, Wolf B, Thalmeier P, Gunther M, Sixl W and Bruls G 1993 Phys. Lett. A {\bf 175} 237

\bibitem{Mydosh:11}
Mydosh J A and Oppeneer P M 2011 Rev. Mod. Phys. {\bf 83}, 1301

\bibitem{Schoenes:87}
Schoenes J, Schonenberger C, Franse J J M and A. A. Menovsky 1987 Phys. Rev. B(R) {\bf 35} 5375

\bibitem{Oh:07}
Oh Y S,  Kim K-H, Sharma P A, Harrison N, Amitsuka H and Mydosh J A 2007 Phys. Rev. Lett. {\bf 98}, 016401

\bibitem{Kasahara:07}
Kasahara Y, Iwasawa T, Shishido H, Shibauchi T, Behnia K, Haga Y, Matsuda T D, Onuki Y, Sigrist M and Matsuda Y 2007 Phys. Rev. Lett. {\bf 99}, 116402

\bibitem{Bonn:88}
Bonn D A, Garrett J D and Timusk T 1988 Phys. Rev. Lett. {\bf 61} 1305

\bibitem{Broholm:91}
Broholm C, Lin H, Matthews P T, Mason T E, Buyers W J L, Collins M F, Menovsky A A, Mydosh J A and Kjems J K 1991 Phys. Rev. B {\bf 43}, 12 809.

\bibitem{Wiebe:07}
Wiebe C R, Janik J A, MacDougall G J, Luke G M, Garrett J D, Zhou H D, Jo Y-J, Balicas L, Qiu Y, Copley J R D, Yamani Z And Buyers W J L 2007 Nature Phys. {\bf 3} 96

\bibitem{Ohkuni:99}
Ohkuni H, Tokiwa Y, Sakurai K, Settai R, Haga T, Yamamoto E, Onuki Y, Yamagami H, Takahashi S and Yanagisawa T 1999 Philos. Mag. B {\bf 79} 1045

\bibitem{Altarawneh:11}
Altarawneh M M, Harrison N, Sebastian S E, Balicas L, Tobash P H, Thompson J D, Ronning F and Bauer E D 2011 Phys. Rev. Let. {\bf 106} 146403

\bibitem{Santander-Syro:09}
A.~F. Santander-Syro, M. Klein, F.~L. Boariu, A. Nuber, P. Lejay, and F. Reinert, Nat. Phys. {\bf 5},  637  (2009).

\bibitem{Schmidt:10}
A.~R. Schmidt, M.~H. Hamidian, P. Wahl, F. Meier, A.~V. Balatsky, J.~D. Garrett, T.~J. Williams, G.~M. Luke, and J.~C. Davis, Nature {\bf 465},  570  (2010).

\bibitem{Aynajian:10}
P. Aynajian, E.~H. da Silva Neto, C.~V. Parker, Y. Huang, A. Pasupathy, J. Mydosh, and A. Yazdani, PNAS {\bf 107},  10383  (2010).

\bibitem{Rodrigo:97}
J.~G. Rodrigo, F. Guinea, S. Vieira, and F.~G. Aliev, Phys. Rev. B {\bf 55}, 14318 (1997).

\bibitem{Okazaki:11}
Okazaki R, Shibauchi T, Shi H J, Haga Y, Matsuda T D, Yamamoto E, Onuki Y, Ikeda H and Matsuda Y 2011 Science {\bf 331} 439

\bibitem{Kernavanois:99}
Kernavanois N, de Rotier P D, Yaouanc A, Sanchez J P, Li K D and Lejay P 1999 Physica B: Condensed Matter {\bf 259} 648

\bibitem{Kuwahara:97}
Kuwahara K, Amitsuka H, Sakakibara T, Suzuki O, Nakamura S, Goto T, Mihalik M, Menovsky A A, de Visser A and Franse J J M 1997 Journal of the Physical Society of Japan {\bf 66} 3251.

\bibitem{Oppeneer:10}
Oppeneer P M, Rusz J, Elgazzar S, Suzuki M-T, Durakiewicz T and Mydosh J A 2010 Phys. Rev. B {\bf 82} 205103

\bibitem{Oppeneer:11}
Oppeneer P M, Elgazzar S, Rusz J, Feng Q, Durakiewicz T and Mydosh J A 2011 Phys. Rev. B {\bf 84} 241102

\bibitem{Das:12}
Das T 2012 Scientific Reports {\bf 2} 596

\bibitem{Fujimoto:11}
Fujimoto S 2011 Phys. Rev. Lett. {\bf 106} 196407

\bibitem{Rau:12}
Rau J G and Kee H-Y 2012 Phys. Rev. B {\bf 85} 245112

\bibitem{Chandra:13}
Chandra P, Coleman P and Flint R 2013 Nature {\bf 493} 621

\bibitem{Altarawneh:12}
Altarawneh M, Harrison N, Li G, Balicas L, Tobash P H, Ronning F and Bauer E D 2012 Phys. Rev. Lett. {\bf 108} 066407

\bibitem{Lynn:2012}
Lynn J W, Chen Y., Chang S, Zhao Y, Chi S, Ratcliff W, Ueland B G and Erwin R W 2012 J. Res. Nat. Inst. Stand. Technol. {\bf 117} 61

\bibitem{Izyumov:91}
Izyumov I A, Naish V E and Ozerov R P 1991 Neutron Diffraction of Magnetic Materials. Plenum Publishing Corporation, New York

\bibitem{Takagi:07}
Takagi S, Ishihara S, Saitoh S, Sasaki H, Tanida H, Yokoyama M and Amitsuka H 2007 J. Phys. Soc. Jpn. {\bf 76} 033708

\bibitem{Takagi:12}
Takagi S, Ishihara S, Yokoyama M And Amitsuka H 2012 J. Phys. Soc. Jpn {\bf 81} 114710


\end{thebibliography}

\end{document}